# Medical Image Analysis for Detection, Treatment and Planning of Disease using Artificial Intelligence Approaches

Nand Lal Yadav[1], Satyendra Singh[1], Rajesh Kumar[1], Sudhakar Singh[1*]

**ABSTRACT**

X-ray is one of the prevalent image modalities for the detection and diagnosis of the human body. X-ray provides an actual anatomical structure of an organ present with disease or absence of disease. Segmentation of disease in chest X-ray images is essential for the diagnosis and treatment. In this paper, a framework for the segmentation of X-ray images using artificial intelligence techniques has been discussed. Here data has been pre-processed and cleaned followed by segmentation using SegNet and Residual Net approaches to X-ray images. Finally, segmentation has been evaluated using well known metrics like Loss, Dice Coefficient, Jaccard Coefficient, Precision, Recall, Binary Accuracy, and Validation Accuracy. The experimental results reveal that the proposed approach performs better in all respect of well-known parameters with 16 batch size and 50 epochs. The value of validation accuracy, precision, and recall of SegNet and Residual Unet models are 0.9815, 0.9699, 0.9574, and 0.9901, 0.9864, 0.9750 respectively.

*Corresponding author

Nand Lal Yadav
nly.jct@gmail.com

Satyendra Singh
satyendra@allduniv.ac.in

Rajesh Kumar
rajeshkumar@allduniv.ac.in

Sudhakar Singh
sudhakar@allduniv.ac.in

[1] Department of Electronics and Communication, University of Allahabad, Prayagraj, India

# 1. INTRODUCTION

Medical Image Analysis (MIA) is a field of study that deals with the processing, analysis, and interpretation of the medical images for diagnosis, treatment planning, and research. It involves the application of various techniques and algorithms to extract meaningful information from medical images, e.g. X-rays, CT scans, MRI scans, and Ultrasound images. The goal of MIA is to design and develop computer-based tools and systems that can assist clinicians in the analysis and interpretation of medical images, improve the accuracy and efficiency of diagnosis and treatment, and enable quantitative assessment of disease progression and treatment outcomes. MIA has widespread applications in various areas of medicine, including radiology, cardiology, neurology, oncology, and ophthalmology [1]. MIA involves a variety of techniques, including image pre-processing, image segmentation, feature extraction, classification, and visualization. The processing steps may involve various types of algorithms, including machine learning, pattern recognition, computer vision, and deep learning. MIA is a rapidly growing field, with new techniques and applications emerging regularly. It has a good potential to significantly improve the accuracy and efficiency of medical diagnosis and treatment and is expected to have a significant impact on the healthcare industry in the coming ages.

Some popular machine learning approaches for the medical image analysis, detection, treatment, and planning are summarized in Table 1 [2], [3]. These machine learning approaches have shown promising results in numerous applications, including disease detection, treatment planning, and prediction. However, it's essential to note that these models should be carefully evaluated and validated before their use in clinical practice.

**Table. 1 Some machine learning approaches for medical image analysis**

| Sl. No. | Machine Learning Models | Usage in Medical Image Analysis |
|---|---|---|
| 1. | Convolutional Neural Networks (CNNs) | Suitable for segmentation, detection, and classification |
| 2. | Recurrent Neural Networks (RNNs) | Suitable for analyzing time-series data, such as medical images captured over time, and for diagnosis and treatment planning |
| 3. | Generative Adversarial Networks (GANs) | Used to generate synthetic medical images that can be used for training and testing deep learning models |
| 4. | Support Vector Machines (SVMs) | Used for image classification tasks, such as identifying cancerous cells or tumors in medical images |
| 5. | Random Forest (RF) | Used for classification tasks in medical imaging, such as differentiating between healthy and diseased tissues |
| 6. | Deep Reinforcement Learning | It involves training deep learning models to interact with medical images to develop a treatment plan, such as determining the optimal radiation dose for a tumor |

The steps involved in medical image processing for disease prediction can vary depending on the specific application and type of disease being diagnosed [4]. However, some general steps typically involved in medical image processing for disease prediction are listed in Table 2.

**Table. 2 Steps in medical image processing for disease prediction**

| Sl. No. | Medical Image Processing Steps | Description |
|---|---|---|
| 1. | Image acquisition | It involves capturing medical images, like X-rays, CT scans, or MRIs. |
| 2. | Pre-processing | It involves pre-processing the medical images to correct for any artifacts or noise that may affect the accuracy of subsequent analysis. Examples of pre-processing techniques include filtering, denoising, and contrast enhancement. |
| 3. | Image segmentation | Medical images are segmented into meaningful regions that can be used for disease detection and diagnosis. This step can be performed manually or using automated segmentation algorithms. |
| 4. | Feature extraction | It involves extracting relevant features from the segmented regions to identify disease-specific patterns. These features can include texture, shape, and intensity measurements. |

| 5. | Feature selection | In this step, the most pertinent features are selected from the extracted features for dimensionality reduction of the data and increasing the accuracy of the disease prediction model. |
|---|---|---|
| 6. | Machine learning model development | It involves developing a machine learning model e.g. a decision tree or support vector machine, that can be trained to identify disease patterns based on the selected features. |
| 7. | Model evaluation | The performance of the machine learning models is evaluated using validation techniques, such as cross-validation or holdout validation. |
| 8. | Disease prediction | Finally, the trained model is used to predict the presence or absence of the disease in new medical images. |

Medical image processing for disease prediction is a complex and iterative process that requires expertise in both medical imaging and machine learning. Proper evaluation and validation of the machine learning model are critical before deploying it in a clinical setting [5].

In this paper, we used X-ray images. Its analysis focuses on the processing and analysis of the X-ray images for the diagnosis and treatment of the various medical conditions. X-rays are a type of electromagnetic radiation that can penetrate through the human body, allowing for the creation of detailed images of the internal structures of the body. In X-ray image analysis, various techniques and algorithms are used to process and analyse the X-ray images to extract useful information that can aid in diagnosis and treatment [6]. Some of the key applications of X-ray image analysis include:

*Detection of abnormalities and diseases:* X-ray image analysis can be applied to detect various abnormalities and diseases in the body, such as fractures, tumours, and pneumonia.

*Quantitative assessment of disease progression:* X-ray images can be used to monitor the progression of certain diseases, such as osteoporosis.

*Treatment planning:* X-ray image analysis can be used to plan surgical procedures and radiation therapy, by providing detailed information about the location and size of tumors or other abnormalities.

Overall, X-ray image analysis is a critical tool for medical professionals and has a wide variety of applications in various areas of medicine, including radiology, cardiology, oncology, and orthopaedics.

In this paper an artificial intelligence (AI) approach for segmentation and detection of Covid-19 from X-ray images has been implemented. For this purpose, firstly data has been cleaned and after cleaning SegNet and ResNet model of AI have been applied to the data. The results have been compared with the existing work with respect to the different metrics for performance evaluation.

The rest of this paper is organised in the following manner. In section 2 various work done in the area of medical image segmentation in the literature have been discussed. Section 3 provides the materials and methods used in this paper and section 4 describes the framework of the proposed approach. Section 5 cover the results and discussions and finally, Section 6 presents the concluding remarks of this paper.

## 2. RELATED WORKS

AI approaches, such as deep learning, can assist in interpretation of the chest X-ray images by automatically detecting and classifying disease patterns. These AI models can analyze large datasets of the chest X-ray images and learn to identify subtle differences between normal and abnormal images, potentially leading to faster and more accurate diagnoses. One of the most successful and effective deep learning models for chest X-ray disease detection is CheXNet, which was developed using a huge dataset of chest X-ray images with associated disease labels. CheXNet can accurately detect a variety of diseases, including pneumonia, tuberculosis, and lung nodules. Other deep learning models, such as DenseNet and ResNet, have also been used for disease detection in chest X-ray with promising results.

Duncan and Ayache [7] examined the advancements in medical image analysis in past two decades and discussed certain important challenges. Chen [8] suggested a CNN model that has undergone training and testing using a publicly available X-ray dataset that was recently released for the classification of tertiary and normal cases. The framework that was suggested shows good performance even when there is a limited amount of data available regarding the pattern of COVID-19.

In the research by Saygılı [9], it was examined that how machine learning enabled image processing can aid in the swift and precise identification of COVID-19 using two of the most commonly used medical imaging methods, namely chest X-rays and CT scans. The study's key objective is to assist medical professionals who are exhausted and under considerable strain during the COVID-19 outbreak by using intelligent learning techniques and image classification methods. The suggested approach achieves high levels of the COVID-19 detection accuracy for three datasets with accuracy rates of 89.41%, 99.02%, and 98.11%, respectively. However, for the X-ray images, the accuracy attained for COVID-19 (+), COVID-19 (-), Pneumonia but not COVID-19 classes is 85.96%.

Kalinovsky and Kovalev [10] carried out research and development on the segmentation of lungs in X-ray images for chest using Deep Learning and Encoder-Decoder Convolutional Neural Networks (ED-CNN). It is an exploratory phase, and the outcomes are reported. This research was conducted in the setting of a population-wide screening for lung and heart illnesses and in the development of computational services for an international portal on lung tuberculosis. The findings suggest that ED-CNN networks have the potential to be a valuable resource for automatically segmenting the lung in large-scale initiatives. In large-scale projects, the use of Encoder-Decoder Deep CNNs is a potentially effective technique for the automatic segmentation of lungs in X-ray images of chest.

Spampinato et al. [11] suggested and experimented with various deep learning techniques to automatically evaluate skeletal bone age. Their findings indicate that the average difference between automatic and manual assessments is around 0.8 years, which is considered as state-of-the-art performance. This study represents the initial attempt to automate skeletal bone age assessment across all age groups, genders, and races, which was evaluated using a public dataset.

Haq et al. [12] suggested a deep learning-based system that utilizes community-based approaches for automatically retrieving medical images from a vast X-ray database. The system is designed to extract images that are similar in nature. In comparison to the most advanced medical image retrieval methods, the proposed approach exhibited superior performance. The suggested technique accomplished an accuracy of 85% in identifying medical images with matching disease labels.

Müller and Kramer [13] developed a Python library called MIScnn, which is available for public use. MIScnn is a system for medical image segmentation that employ CNNs and deep learning techniques. To prove the effectiveness of MIScnn, authors conducted an automatic cross-validation using the Kidney Tumor Segmentation Challenge 2019 CT dataset, which resulted in a robust predictor.

Girum et al. [14] introduced a novel, effective deep learning model for precisely segmenting targets from the images, while concurrently producing an annotated dataset that can train other deep learning models. The method comprises the utilization of a generative neural network to predict prior knowledge from pseudo-contour landmarks. Authors suggested a swift, responsive deep learning technique to achieve precise segmentation of medical images.

Huang et al. [15] described a fresh segmentation structure that incorporates the anatomical prior of medical images through loss and integrates it into deep learning models. The framework consists of crucial information such as the prior knowledge of the location and shape of organs, that is imperative for precise segmentation. Additionally, they merged the newly proposed deep atlas prior loss with the traditional likelihood losses like Dice loss and focal loss in a Bayesian framework. This framework is adaptive and comprises a prior and likelihood, resulting in an adaptive Bayesian loss. Although the technique outperforms others, it is essential to recognize its limitations. The probabilistic atlas is effective for organs with a relatively consistent position but becomes less useful for tumors with significant positional and morphological changes.

Alom et al. [16] suggested a model that includes two neural network models, namely the Recurrent Convolutional Neural Network (RCNN) based on U-Net, and the Recurrent Residual Convolutional Neural Network (RRCNN) based on U-Net, which we refer to as RU-Net and R2U-Net, respectively. These models integrate the abilities of U-Net, Residual Network, and RCNN to enhance their performance. The suggested models used the three standard datasets for skin cancer segmentation, lung lesion segmentation, and blood vessel segmentation in retina images. The experimental results demonstrated that their models outperform equivalent models, such as UNet and residual U-Net (ResU-Net), on segmentation tasks.

## 3. MATERIALS AND METHODS

The materials and methods used for disease detection in X-ray images of chest using AI approaches can vary depending on the specific study, but here are some common elements.

### 3.1 Dataset

A large image dataset of chest X-ray is required for training and validating the AI model. Typically, this dataset will be labeled with the presence or absence of specific diseases, such as pneumonia, tuberculosis, or lung cancer. The source of the dataset used in this paper is "National Library of Medicine, National Institutes of Health, Bethesda, MD, USA and Shenzhen No. 3 People's Hospital, Guangdong Medical College, Shenzhen, China" [17], [18]. The dataset consists of the X-ray images and the corresponding segmented masks.

### 3.2 Methodology

In our proposed approach X-ray image segmentation has been used for analysis of lung disease. X-ray image segmentation is the process of separating the regions of an X-ray image that correspond to different anatomical structures. Segmentation is a crucial step in many medical imaging applications, including diagnosis, treatment planning, and image-guided interventions. There are several approaches to X-ray image segmentation listed as follows.

*Thresholding:* It is a simple image segmentation approach that involves separating an image into two classes based on a threshold value. This approach can be effective for segmenting X-ray images with clear boundaries between the regions of interest and the background.

*Region-based segmentation:* This approach involves grouping similar pixels in an image based on their intensity or texture properties. The most common region-based approach is the watershed algorithm, which segments an image based on the local minima of the image's gradient magnitude.

*Edge-based segmentation:* This approach involves identifying the edges in an image and then separating the regions based on these edges. Edge-based segmentation is commonly used for X-ray image segmentation, particularly when there is a clear boundary between the structures of interest.

*Deformable models:* Deformable models are a class of segmentation approaches that use mathematical models to represent the boundaries of anatomical structures. The most common deformable models used in X-ray image segmentation are active contours or snakes, which deform to fit the edges of the structures.

*Deep learning-based segmentation:* Recently, deep learning-based segmentation approaches have been used for X-ray image segmentation [19]. These approaches use CNNs to learn the features and boundaries of the structures of interest from a large dataset of annotated X-ray images.

The above X-ray image segmentation approaches have their strengths and weaknesses and are selected based on the specific imaging modality and application requirements. Each approach requires careful tuning of parameters and validation to ensure accurate segmentation.

Chest X-ray is a very popular imaging modality used for the detection and classification of respiratory diseases such as pneumonia, tuberculosis, and lung cancer. The detection and classification of these diseases in the chest X-ray images involve several steps [20]–[22] described in Table 3.

**Table. 3 Steps in detection and classification of disease in chest X-ray image**

| Sl. No. | Medical Image Processing Steps | Description |
|---|---|---|
| 1. | Image pre-processing | This step involves pre-processing the chest X-ray images to enhance the contrast and remove any noise that may interfere with disease detection. |
| 2. | Region of Interest Identification | In this step, the regions of interest (ROI), such as lung fields, are identified and extracted from the chest X-ray image. |
| 3. | Image segmentation | The regions of interest are segmented into meaningful regions based on their intensity or texture properties. |
| 4. | Feature extraction | Relevant features such as texture, shape, and intensity measurements are extracted away from the segmented regions. |
| 5. | Classification | The extracted features are used to train a model, such as an SVM or CNN or deep learning, to classify the X-ray images of chest into different disease categories. |
| 6. | Validation | Performance of the machine learning models are validated using various techniques such as cross-validation or holdout validation. |
| 7. | Disease detection | The trained model is used to detect the existence or absence of the disease in new chest X-ray images. |

Several deep learning methods have been applied successfully for disease detection in chest X-ray images, particularly for the detection of respiratory diseases such as pneumonia, tuberculosis, and lung cancer [23]. Some popular deep learning models for disease detection in chest X-ray images are as follows.

*CheXNet:* It is a CNN model trained on a large dataset of chest X-ray images with associated labels for fourteen different pathologies. It can accurately detect diseases such as pneumonia, atelectasis, and cardiomegaly [24].

*Deep Chest:* It is another CNN-based deep learning model trained on a dataset of over 70,000 chest X-ray images. It can detect multiple pathologies, including pneumonia, tuberculosis, and lung nodules, with high accuracy.

*DenseNet:* It is a deep learning model that uses densely connected convolutional layers to extract the features from chest X-ray images. It has been used for the detection of lung nodules and has shown promising results in the detection of lung cancer.

*ResNet:* It is another popular deep learning model that has been used for the detection of various respiratory diseases in chest X-ray images. It uses residual connections to improve the training of deep neural networks and has shown the improvement in accuracy of disease detection [25].

*Inception-ResNet-v2:* It is a contemporary deep learning model that has been used for the detection of tuberculosis in chest X-ray images. It combines the Inception and ResNet architectures to improve the accuracy of disease detection.

The accuracy of the disease detection and classification models varies depending on the X-ray images quality, the complexity of the disease, and the size of the dataset used for training. Proper evaluation and validation of the machine learning model are critical before deploying it in a clinical setting.

## 4. PROPOSED APPROACH

SegNet is a deep learning model that has been primarily used for image segmentation tasks. It is based upon a fully convolutional neural network architecture, with an encoder-decoder network that learns to map an input image to an output segmentation map [26]. While SegNet was not specifically designed for disease detection in chest X-ray images, it can be adapted for segmentation task [27], [28].

To use SegNet for chest X-ray disease detection, the model can be trained on a chest X-rays image dataset with associated disease labels. The encoder network of SegNet learns to extract relevant features/fields from the X-ray images of chest, while the decoder network performs the segmentation task by generating a segmentation map that highlights areas of the image that are indicative of disease [29]. One potential advantage of using SegNet for detection of chest X-ray disease is that it can identify multiple diseases in a single image. For example, if a chest X-ray image contains both pneumonia and a lung nodule, SegNet can be trained to segment both pathologies in the same image.

While SegNet has shown promising results for medical image analysis tasks including disease detection in X-ray images of chest, it is not as widely used as other deep learning models such as CheXNet and Deep Chest. It is important to mention that the accuracy of SegNet for disease detection in X-ray images of chest will depend on the quality and size of the dataset used for training as well as on the complexity of the diseases being detected. Proper evaluation and validation of the model is critical before deploying it in a clinical setting.

Clinical correlation is an important aspect of disease detection in X-rays images of chest using any deep learning model including SegNet. While SegNet can accurately identify regions of the chest X-ray image that are indicative of a disease, it is ultimately up to a medical professional to interpret these results in the context of the patient's clinical presentation [30].

In other words, the SegNet model can assist radiologists and other medical professionals in identifying potential areas of concern in a chest X-ray image, but it cannot make a diagnosis on its own. The medical professional must use their clinical knowledge and expertise to interpret the SegNet results, considering the patient's medical history, symptoms, and other diagnostic tests. In Fig. 1, the framework of the proposed approach for X-ray image segmentation is illustrated.

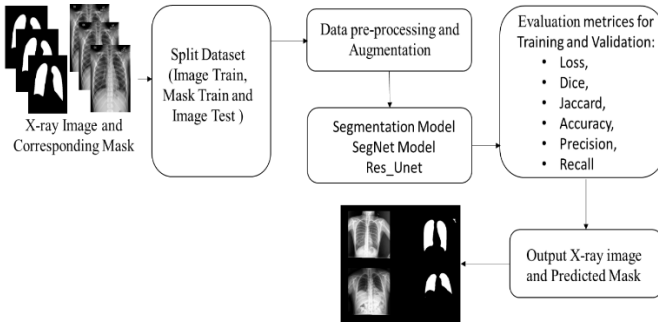

**Fig. 1** Illustration of the proposed approach

For example, if the SegNet model identifies an area of the chest X-ray image as potentially indicative of pneumonia, the radiologist must determine if this finding is consistent with the patient's clinical presentation, including their symptoms, physical exam findings, and other diagnostic tests. If there is a clinical correlation between the SegNet results and the patient's presentation, further testing or treatment may be warranted. However, if there is no clinical correlation, the SegNet finding may be a false positive, and further investigation may not be necessary. Overall, the clinical correlation of any deep learning model for X-ray image of chest disease detection is critical for ensuring accurate and appropriate diagnosis and treatment of patients. The SegNet model involves several steps which include the following.

*Data pre-processing:* The chest X-ray images must be pre-processed to ensure that they are in a suitable format for input to the SegNet model. This may involve resizing the images, normalizing pixel values, and converting the images to grayscale.

*Training data preparation:* Annotated training data must be prepared, which involves manually segmenting the chest X-ray images to identify areas of interest that correspond to various diseases. This step requires medical expertise and is resource and time consuming.

*SegNet model architecture:* The SegNet model is based on a fully convolutional neural network architecture, with an encoder-decoder network that learns to map an input image to an output segmentation map. The encoder network learns to extract relevant features/fields from the X-ray images of chest, while the decoder network performs the segmentation task by generating a segmentation map that highlights areas in the image that are indicative of a particular disease.

*Training the SegNet model:* The SegNet model is trained on annotated training data, using backpropagation to adjust the weights in the network based on the difference between the predicted segmentation map and the ground truth segmentation map.

*Validation and testing:* The SegNet model is evaluated on a validation dataset to ensure that it is not overfitting to the training data. Once the model has been validated, it can be tested on a separate test dataset to evaluate its performance.

*Clinical correlation:* As noted earlier, the final step in the SegNet model involves clinical correlation, where a medical professional uses the SegNet results in the context of the patient's clinical presentation to make a diagnosis.

Overall, the SegNet model is an efficient and powerful tool for disease detection in chest X-ray images, but it requires significant time and expertise to prepare and train the model properly. Proper evaluation and validation of the model is critical before deploying it in a clinical setting.

Here are several performance evaluation metrics [31], [32] that can be used to assess the performance of the machine learning models for classification of chest x-ray image.

*Accuracy:* It is the percentage of samples correctly classified out of all samples.

Accuracy = $(Tp+Tn)/(Tp+Tn+Fp+Fn)$ (1)

*Precision:* It is the percentage of true positive samples out of all predicted positive samples.

Precision = $Tp/(Tp+Fp)$ (2)

*Recall:* It is the percentage of true positive samples out of all actual positive samples.

Recall = $Tp/(Tp+Fn)$ (3)

*F1 score:* It is the harmonic mean of the precision and recall.

F1 score = 2(precision * recall)/(precision + recall) (4)

where $Tp$ is true positive, $Tn$ is true negative, $Fp$ is false positive, and $Fn$ is false negative. Equations (1), (2), (3), and (4) show the evaluation metrics accuracy, precision, recall, and F1 score respectively.

*The area under receiver operating characteristic curve (AUC-ROC):* It is the measure ability of model to discriminate between positive and negative samples.

*Confusion matrix:* It is a matrix that summarizes that how many true positives, true negatives, false positives, and false negatives are.

*Sensitivity:* It is the percentage of the true positive samples out of all actual positive samples.

*Specificity:* It is the percentage of the true negative samples out of all actual negative samples.

The choice of evaluation metrics depends on the specific classification problem and the desired performance criteria.

## 5. EXPERIMENTAL RESULTS AND DISCUSSION

Experimental results for SegNet and Residual Unet based X-ray

image segmentation models have shown that it is effective for semantic segmentation tasks. The result of our proposed approach is shown in Table 4 along with a comparison with the one proposed by Chavan et al. [26].

**Table. 4 Comparison of proposed work with existing work**

|  | Model | Loss | Dice Coef | Specificity | Sensitivity | Recall | Precision |
|---|---|---|---|---|---|---|---|
| **Chavan et al. [30]** | UNet Train | 0.3216 | 0.6785 | 0.9822 | 0.9779 | 0.9776 | 0.9512 |
|  | UNet Validation | 0.3232 | 0.6775 | 0.9797 | 0.9719 | 0.9711 | 0.9439 |
| **Proposed Approach** | Segnet Train | 0.0329 | 0.9671 | -- | -- | 0.9621 | 0.9723 |
|  | Segnet Validation | 0.0363 | 0.9636 | -- | -- | 0.9574 | 0.9699 |
|  | Res_unet Train | 0.0127 | 0.9873 | -- | -- | 0.9868 | 0.9879 |
|  | Res_unet Validation | 0.0193 | 0.9808 | -- | -- | 0.9750 | 0.9864 |

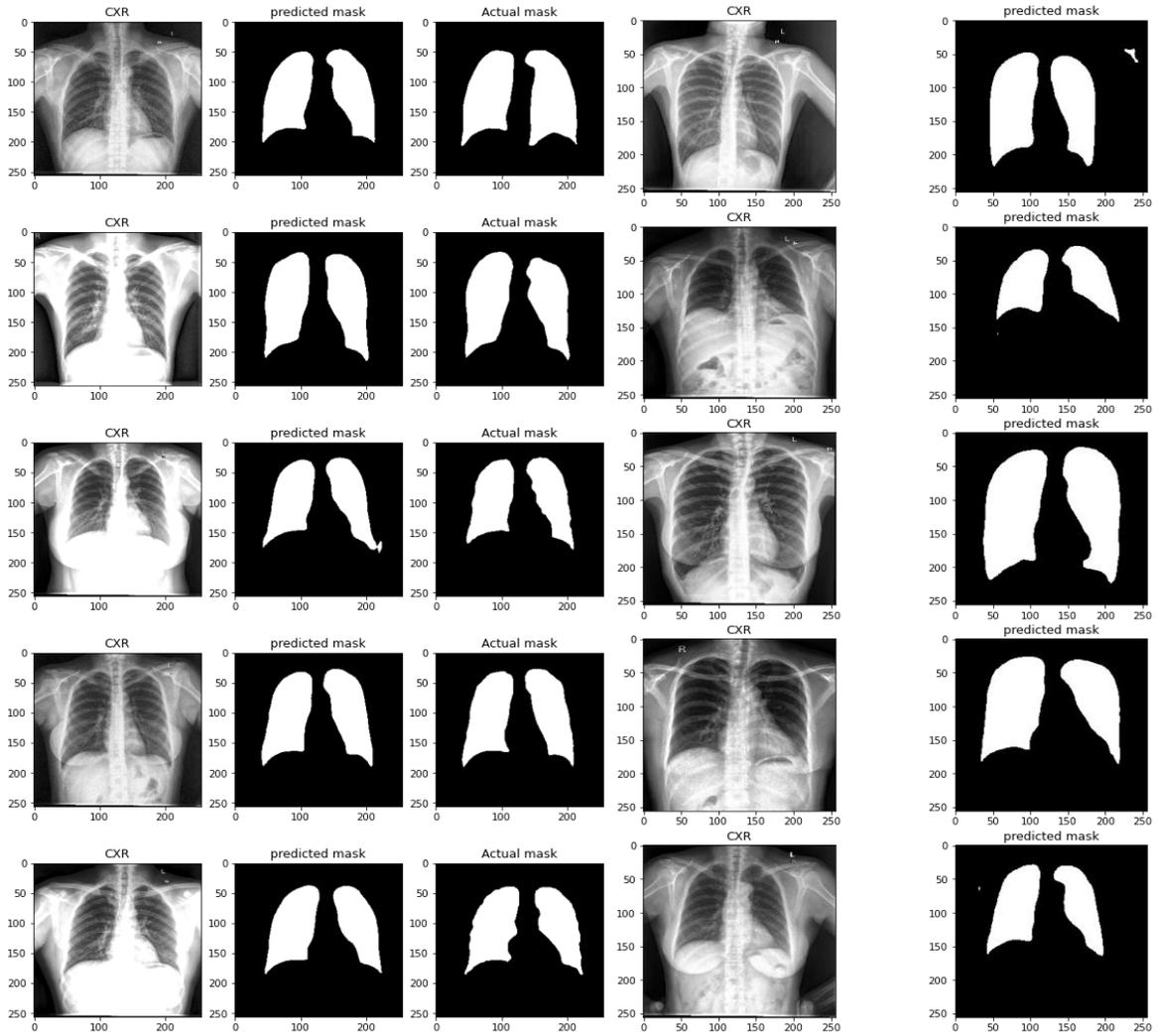

**Fig. 2** Actual X-ray image, Predictive mask, and Actual Mask and right side of the figure X-ray image and predictive mask

From Table 4, it can be seen that the Residual Unet model performs better in comparison to the SegNet and Unet models. Fig. 2 visualises the result of the proposed deep learning based X-ray image segmentation approach. The resulting images are X-ray image, mask image, and predicted mask image.

The experimental results have shown that SegNet is more effective than other considered models. In this work, obtained metrics are loss, dice coefficient, Jaccard coefficient, binary accuracy, precision, recall, validation accuracy, validation recall, validation precision, validation loss, validation dice coefficient, and validation Jaccard coefficient. The curves that demonstrate the cumulative frequencies of data for both Dice and Jaccard methods are known as Cumulative Frequency Curves for Dice and Jaccard. In the proposed work, we have 16 batch size and 50 epochs for training the model. The value of different metrics of Segnet and Residual Unet models are shown in Table 5.

Figure 3 shows the Cumulative Frequency Curves of the SegNet model that involves Dice and Jaccard coefficients.

**Table. 5** Value of different metrics of Segnet and Residual Unet models

| Models ⇒ <br> Metrics ⇓ | Segnet Model | Residual Unet Model |
| --- | --- | --- |
| loss | 0.0329 | 0.0127 |
| dice_coef | 0.9671 | 0.9873 |
| jaccard_coef | 0.9363 | 0.9749 |
| binary_accuracy | 0.9835 | 0.9936 |
| precision | 0.9723 | 0.9879 |
| recall | 0.9621 | 0.9868 |
| val_loss | 0.0363 | 0.0193 |
| val_dice_coef | 0.9636 | 0.9808 |
| val_jaccard_coef | 0.9299 | 0.9624 |
| val_binary_accuracy | 0.9815 | 0.9901 |
| val_precision | 0.9699 | 0.9864 |
| val_recall | 0.9574 | 0.9750 |

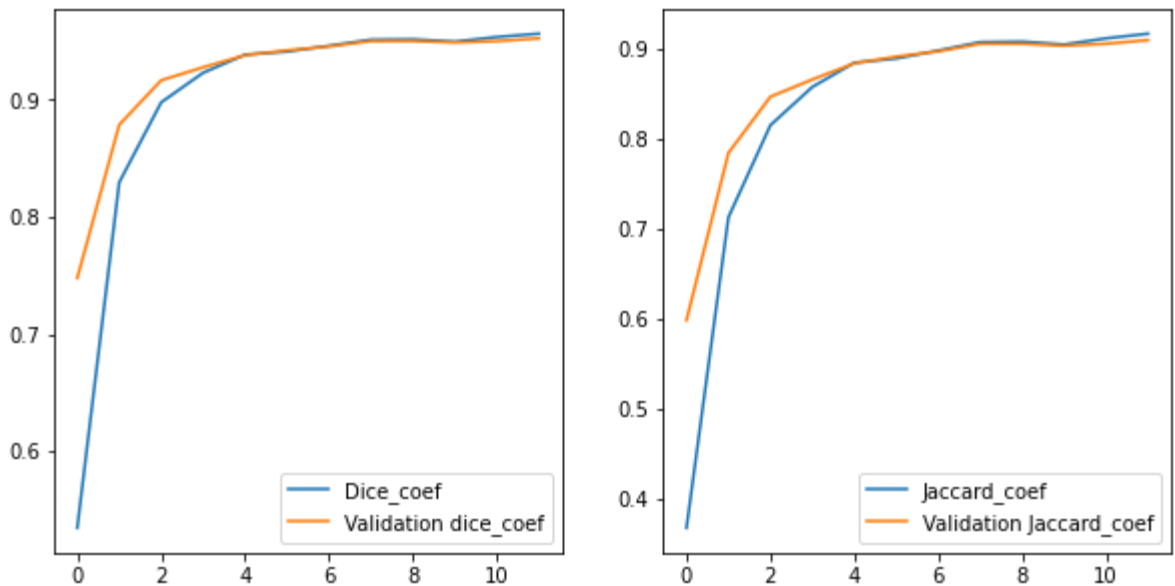

**Fig. 3** Epoch vs Value of Dice coefficient, Validation dice coefficient, and Jaccard coefficient, Validation Jaccard coefficient

## 6. CONCLUSION

The use of AI approaches for disease detection in chest X-ray has the potential to revolutionize medical imaging and improve patient care. AI models can support in the early detection of diseases, leading to the earlier interventions and improved outcomes. This paper uses a deep learning-based X-ray image segmentation model with the help of SegNet and Residual Unet model. From the obtained experimental results, it is observed that the performance of the proposed deep learning approach is better in comparison to other existing models. This model is helpful for the early detection of lung disease. However, it is important to note that AI approaches are not a substitute for medical expertise, and the final diagnosis should always be made by a medical professional in the context of the patient's clinical correlation. In the future, we extend this model for multilevel segmentation of X-ray images.